 \documentclass[preprint2]{aastex}

\input{psfig.sty}

\shorttitle{Scientific impact of large telescopes}
\shortauthors{Benn \& S\`{a}nchez}

\begin{document}


\title{Scientific impact of large telescopes}

\slugcomment{Accepted for publishing in PASP}

\author{C.R. Benn and S.F. S\'{a}nchez}
\affil{Isaac Newton Group, Apartado 321, 38700 Santa Cruz de La Palma,
Spain}
\email{crb@ing.iac.es}




\begin{abstract}
The scientific impacts of telescopes worldwide have been compared
on the basis of their contributions to (a) the 
1000 most-cited astronomy papers published 1991-8
(125 from each year), and (b) the 452 astronomy papers 
published in Nature
1989-98.
1-m and 2-m ground-based telescopes account for
$\approx$ 5\% of the citations to the top-cited papers,
4-m telescopes 10\%, Keck I/II 4\%, sub-mm and radio
telescopes 4\%, HST 8\%, other space telescopes 23\%.
The remaining citations are mainly to theoretical and review papers.
The strong showing by 1-m and 2-m telescopes in the 1990s
augurs well for the continued scientific impact
of 4-m telescopes in the era of 8-m telescopes.
The impact of individual ground-based optical telescopes 
is proportional to
collecting area (and approximately proportional to capital cost).
The impacts of the various 4-m telescopes are
similar, with CFHT leading in citation counts, and WHT in Nature papers.
HST has about 15 times the citation impact of a 
4-m ground-based telescope, but cost $>$ 100 times as much.
Citation counts are proportional to counts of papers published in Nature,
but for radio telescopes the ratio is a factor $\sim$ 3  smaller 
than for optical telescopes,
highlighting the danger of using either metric alone to compare
the impacts of different types of telescope.
Breakdowns of citation counts by
subject (52\% extragalactic),
and journal (ApJ 44\%, Nature 11\%, MNRAS 9\%, A\&A 6\%)
are also presented.

\end{abstract}


\keywords{telescopes}



\section{Introduction}
New telescopes are funded on the basis of
the expected yield in terms of new science,
so it is of interest to compare the current scientific
impact of different kinds of telescope,
e.g. small vs large, optical vs radio,
ground-based vs space.
Scientific impact can be measured in a number of ways,
including 
(1) counting published papers, 
(2) counting publications in high-impact journals, 
and 
(3) counting citations to published
papers.
(1) Counts of papers are easy to make, and several observatories maintain
lists of papers based on data obtained with 
their telescopes, but such counts give equal credit to papers
of different scientific merit.
(2) Alternatively, one can look at the origin of
only the highest-cited papers, or of papers appearing in journals
with the highest impact, but the
statistical noise is then higher.  
Lamb (1990) used this approach to chart the changing
strength of biological research in different countries.
(3) Citation counts provide a fairer measure of the amount of
interest generated by papers, and are often used to assess
the performance of individuals and organisations, 
but suffer from a number of biases:
\newline 
$\bullet$
Geographical bias arises from the tendency of each community to
over-cite its own results e.g. through 
reading and citing national journals (Abt \& Liu 1989, 
Schoonbaert \& Roelants 1998), or e.g. through
prefentially citing results first heard at regional conferences.
\newline
$\bullet$ 
Language bias arises from the tendency of
English-speaking scientists not to read or
cite papers written in other languages (noted e.g. by Rees 1997)
and, more subtly, from
citation errors (Abt 1992,
Kotiaho et al 1999) which may cause non-English citations to be missed
more often by automatic searches.
\newline
$\bullet$ 
Different scientific disciplines cite at different rates
(Abt 1987), 
and this may be true of different sub-disciplines within
astronomy as well. 
Also, 
minority fields may be under-cited,
and popular topics over-cited, relative to real scientific impact.
\newline
$\bullet$
High citation rates for
technical papers may reflect usefulness rather than scientific
impact.
\newline
$\bullet$ Self-citations (Trimble 1986) are included in citation databases,
although for highly-cited papers they are only a small fraction of
the total.
Mutual-citation networks (Coghlan 1991) can also bias
the statistics.
\newline
Another disadvantage 
of citation statistics is the long time lag between data
being taken at the telescope, and citations being made.
Citation rates peak about 5 years after publication,
according to Abt (1998), and it might take much longer for the
significance of 
seminal results to be widely recognised.

Finally, scientific impact is not the only measure of success
for a telescope.  Continued funding, whether by the taxpayer
or from private sources, depends in part on
public interest in individual telescopes,
as reflected e.g. in articles in popular journals.

\subsection{Previous studies}
Abt (1985) compared counts of citations of papers generated by 
two telescopes run by national observatories (the CTIO and KPNO 4-m)
with those from two telescopes run by private observatories
(the Lick 3-m and Palomar 5-m), and found them to be statistically
indistinguishable.

Trimble (1995) compared the impacts of large US optical
telescopes, using the 1993 citation rates of papers published 
in 1990-91, and extended this analysis in Trimble (1996) to include
29 telescopes worldwide.
The telescopes yielding the most citations were,
in decreasing order, the AAT 3.9-m, CTIO 4-m, KPNO 4-m, CFHT 3.6-m, 
Palomar 5-m
and Isaac Newton 2.5-m (see Table 4 for a list of abbreviations), 
with a factor of 2 between the first and sixth ranked.

Leverington (1996) looked at samples of the 15\% most-cited papers in ApJ
and MNRAS taken every 4 years between 1958 - 1994, and found the
six most productive ground-based optical 
telescopes to be the
Palomar 5-m, AAT, Lick 3-m, KPNO 4-m, CTIO 4-m and Palomar Schmidt.
The VLA, and the Einstein, IUE, CGRO and ROSAT satellites 
were comparably productive.
He found that paper counts for space telescopes 
peak about 4 years 
after launch, and fall to half that rate about 9 years after 
launch.
In Leverington (1997a), he showed that 4-m-class ground-based telescopes 
were about 50\% more cost-effective 
(in terms of cost per paper) than 2-m-class telescopes.
Leverington (1997b) extended this analysis to radio and space
telescopes, finding that ground-based optical telescopes were (up to 1994) 
2 - 3 times as cost-effective as space telescopes.  
HST appears particularly inefficient when judged by this metric,
but as Pasachoff (1997) pointed out, this may highlight
the limitations of paper-counting rather than indicate a poor
return on investment.

Gopal-Krishna \& Barve (1998) analysed astronomy papers published 
in Nature 1993-5 and found that of the 51 based on data from ground-based
optical telescopes, 45\% used only data from telescopes with mirror
diameter $<$ 2.5 m.
They concluded that much front-rank science is still done by modest-sized
telescopes, and that the productive life of optical telescopes is
long.

Bergeron \& Grothkopf (1999) ranked selected
telescopes by publication counts, and found the order for 1997/8 to be
WHT, ESO 3.6-m, NTT, AAT, ESO 2.2-m, INT, CFHT, CTIO 4-m, KPNO 4-m. 

Massey et al (2000) used counts of papers from the KPNO 4-m and WIYN 
telescopes to compare the scientific impact of classical and queue observing,
and 
found no evidence that the latter has higher impact.

Citation analyses of the productivity of US, UK and
Dutch astronomers of different standing were made 
by Trimble (1993a), Trimble (1993b) and Spruit (1996) respectively.

\subsection{This paper}
We made two independent analyses of the scientific impact
of telescopes of different sizes and types.
(1) For a citations-based metric, we drew on a list of the
125 most-cited astronomy papers published in each year
1991-8 (1000 papers in all),
calculating the fraction of citations attributable to papers based
on data from each telescope.
(2) For a metric which is free of some of the biases affecting citation
counts, and which responds faster to scientific trends
(but which may still be affected by regional and subject biases),
we analysed counts of papers 
in the journal Nature, to which much of the
best science is submitted.
Nature has the highest impact factor (citations/article) of any journal,
about 8 times greater than that of MNRAS or ApJ.
In addition, for a metric reflecting public interest, we counted mentions 
in the journal Sky \& Telescope.

Throughout, we use the word `impact' (unless otherwise qualified)
to refer to the normalised fraction of citations to the 1000
most-cited papers.

\begin{figure}[t]
\centerline{\vbox to 7.1cm{
\psfig{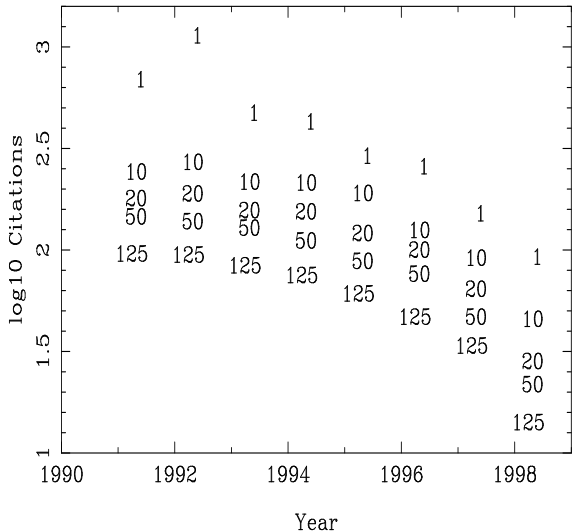}
}}
\caption{Number of citations, up to June 1999, for the 1st, 10th,
20th, 50th and 125th ranked papers published in each year.}
\label{fig:fig1}
\end{figure}

\section{Data}
A list of the 125 most-cited astronomy/space
papers for each year 1991-8, 1000 papers 
in all, was obtained from
David Pendlebury of the Institute for Scientific Information (ISI)
in Philadelphia.

The ISI list covers 21 journals, including
A\&A, A\&AS, ApJ, ApJS, AJ, ARA\&A, MNRAS, Nature, PASP, Science, and
Solar Physics
(using the standard abbreviations).  
For each paper the list gives
the name of the first author, an abbreviated title, 
and the count of citations to that paper up to June 1999.
We added the names of the telescopes
which furnished the data on which each paper was based, 
and a subject code:
sun, solar system, exoplanets, cool stars,
hot stars (including gamma-ray bursts), 
the Galaxy,
galaxies, active galaxies or cosmology.
450 of the 1000 papers were not classified by telescope,
being theoretical (27\%), or reviews (6\%) or using data from
many ($>$ 5) telescopes 
(8\%) or telescopes which could not be identified (5\%).
The fraction of theoretical papers is similar to that found by Abt (1993)
for US astronomical journals as a whole, 28\%. 
Citations of reviews comprise about 1\% of all citations (Abt 1995).
\begin{figure*}[t]
\centerline{\vbox to 11cm{
\psfig{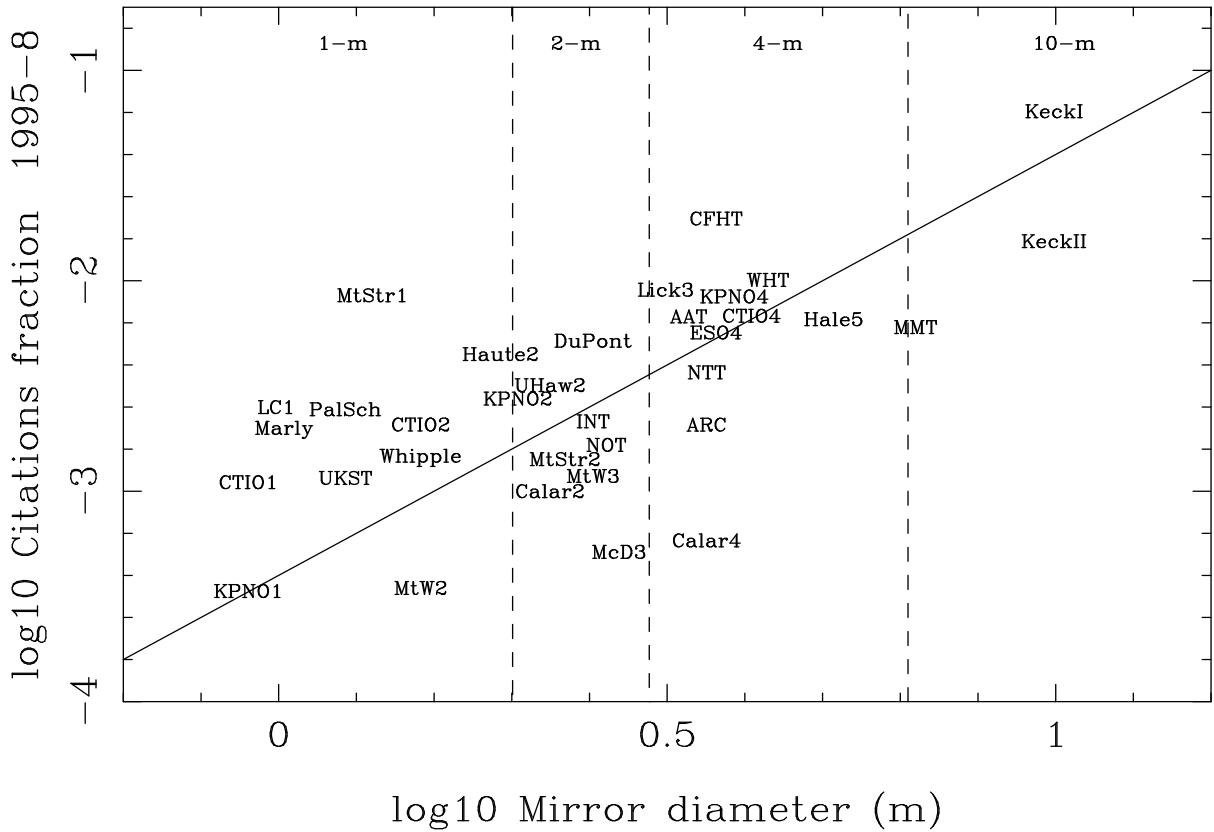}
}}
\caption{Citations fraction 1995-8 vs telescope diameter,
for ground-based optical telescopes.
The straight line corresponds to citation fraction =
0.6\% *(diameter/4 m)$^2$.
Statistical errors $N_{papers}^{-0.5}$ are $\sim$ 0.2 in log10 
for typical
4-m telescopes, $\sim$ 0.3 for 2-m.
In this figure, and others in this paper, a few points have been
displaced slightly to avoid overlap of labels.
}
\label{fig:fig2}
\end{figure*}

The number of citations decreases with increasing year
of publication (Fig. 1, see also
Abt 1981), e.g.
the 20th ranked paper of 1998 has only 26 citations, whereas the
20th ranked in 1991 has 158.
In order to
give equal weight to similarly-ranked papers published in different years,
we expressed the citation count for each paper as a fraction of the
total of citations for the 125 top papers published that year.
With this normalisation, the highest-cited paper of 1991-8
is that by Smoot et al (1992) (COBE results).

For each telescope, we summed the number of papers, and calculated
the average  
citation fraction, for the periods 1991-4, 1995-8 and 1991-8.
About 1/3 of the observational papers are based on data from more
than one telescope (this figure rises to 2/3 for papers based on
data from 4-m telescopes).
In each of these cases, 
the count of papers for each telescope $i$ was incremented by
a weight $w_i$ where $\Sigma w_i$ = 1,
and where the relative weights used for different types of telescope
were 1 for
1-m ground-based optical/IR, 4 for
2-m, 10 for 4-m, 50 for 10-m, 10 for sub-mm, 10 for radio, 
100 for HST, and 30 for other space telescopes.
E.g. if a paper used data from a 4-m class telescope and from ASCA
(space telescope),
the count of papers for the 4-m telescope was incremented by 0.25,
that of ASCA by 0.75
(and similarly for the citation count).
The relative weights of optical ground-based telescopes were estimated
assuming that scientific impact $\propto$ collecting area,
justified by examination of Fig. 2.  
The relative weights of 
ground-based optical and other telescopes were based on 
average impacts deduced from Fig. 3.
Initially, the counts were calculated with all weights set to 1.
The weights were then estimated from Figs. 2 and 3 and the analysis repeated,
this time with weighting.
The new versions of Figs. 2 and 3 were then found to be 
consistent with the applied weighting.
It should be emphasised that 
that changing all the weights to 1 significantly changes
the counts for only a few small telescopes.
The weights were used principally to avoid over-estimation of the 
contribution from 1-m telescopes, since these are often
cited in conjunction with larger telescopes.
The other conclusions of this paper are not affected by the
choice of weighting.

\begin{figure*}[t]
\centerline{\vbox to 11cm{
\psfig{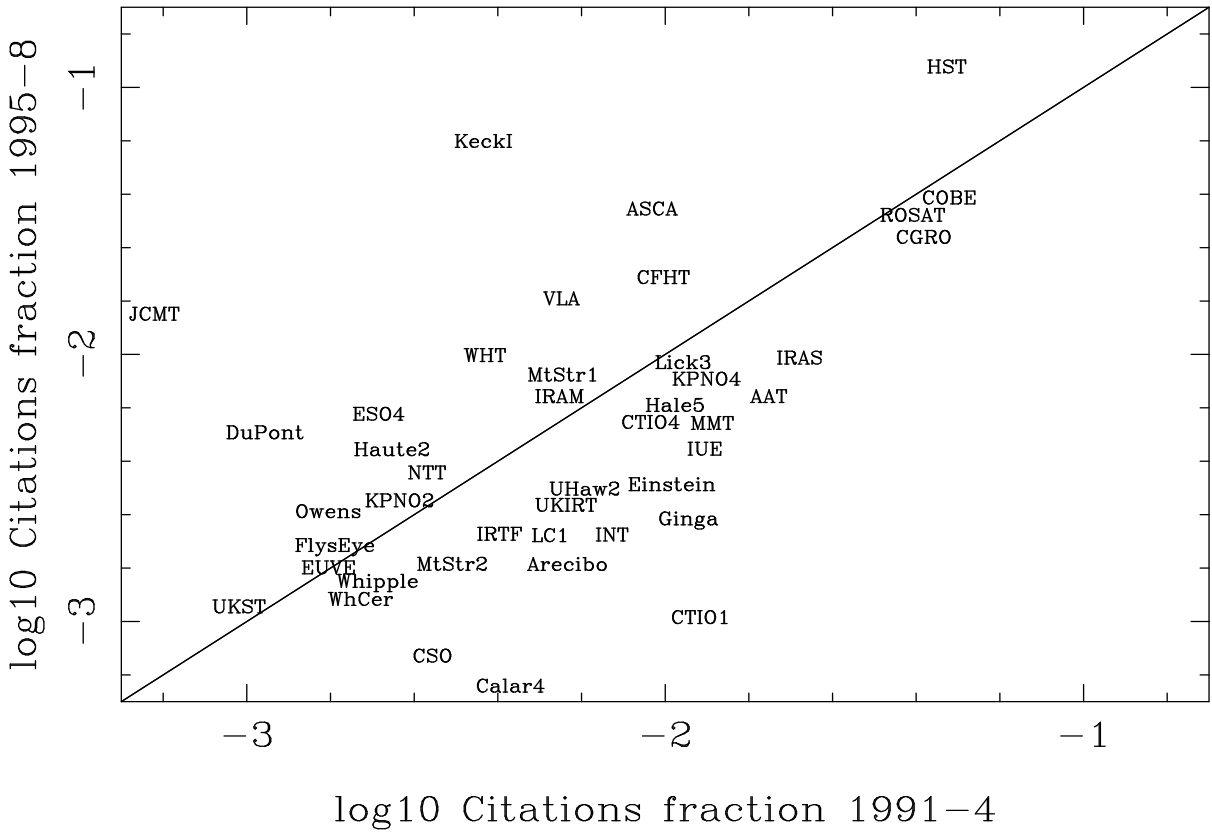}
}}
\caption{Citations fraction 1995-8 vs 1991-4, all telescopes.
The straight line has slope 1.
The telescopes with no citations 1991-4, 
but log10 citations fraction (1995-8)
$>$ -2 are BeppoSAX, Hipparcos, ISO, Keck II, RXTE, and SOHO. 
}
\label{fig:fig3}
\end{figure*}

For none of the large ground-based or 
space telescopes are the counts dominated by
citations to only one or two papers.
Nor are the citation statistics much affected by technical
papers and catalogues receiving unusually large numbers of citations.
A rare example is
Landolt's (1992) paper on photometric standards, 
the second most-cited paper of that year,
which accounts for the high 1991-4 impact of the CTIO 1-m.

Impacts as measured by counts of papers in the ISI
database, and counts of citations to those papers, are similar,
since there is only a weak dependence of citation count on rank
in the ISI database (e.g. a factor of two between 20th and 125th ranked).

We made a similar analysis of the 452 
observational astronomy/space papers
(including letters, articles and reviews) 
published in Nature during
the 10 years 1989-98, again distributing counts 
between telescopes where more
than one contributed to a given paper.
The number of astronomy papers published in Nature each year 
did not change significantly during this period.
The slightly earlier start date than for the citations analysis
allows coverage of a similar sample of new science, since
publication in Nature is rapid.
Approximately 100 of the Nature papers appear also in the sample of
the most-cited 1000 papers.

Finally, in order to gauge public interest in different telescopes,
we counted mentions in the `News Notes' 
section of Sky \& Telescope during 1989-98.

\section{Results}
For each telescope, counts of papers from the 1000 most-cited, 
average citation fractions (to these papers) for 1991-4 and
1995-8, and counts of papers in Nature and of mentions
in Sky \& Telescope are given 
in Tables 1 (ground-based optical),
2 (other ground-based) and 3 (space).
The tables include all
telescopes with a total of more than 1.0 paper
in the top-cited 1000, or in the Nature sample, 
or generating more than 0.1\% of citations in either period.
For reference, Table 1 also lists
all ground-based optical (non-military)
telescopes with primary-mirror
diameter $>$ 3.5 m which had been commissioned by 1996.
The numbers of theoretical and review papers (and citations to them)
are included at the bottom of Table 1.
The counts in Tables 1 - 3 are sums of fractions (as discussed 
in Section 2 above).
Fractional
statistical errors on the telescope 
citation counts are thus $< 1/\sqrt{N}$,
where $N$ is the number of papers to which citations are made.

\begin{figure*}[t]
\centerline{\vbox to 7cm{
\psfig{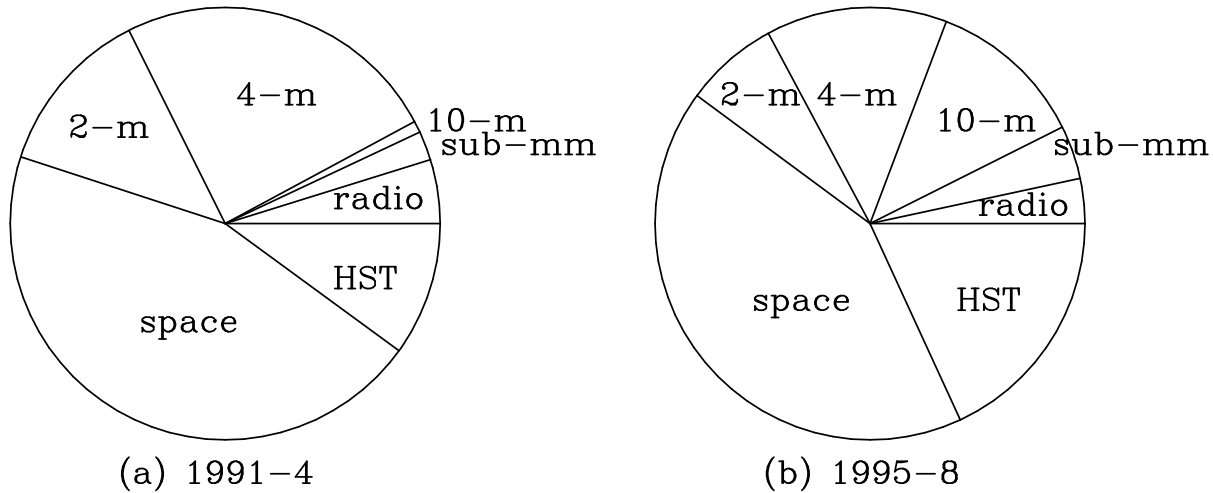}
}}
\caption{Shares of citations
for telescopes of different types, 
for 1991-4 and 1995-8.
`2-m' in this figure includes both 1-m and 2-m class telescopes.
The area of each Venn diagram represents the citations to the 55\% of 
the 1000 papers
for which contributing telescopes could be identified
(i.e. excluding review papers etc.).
The 1991-8 averages are: 1-m + 2-m 9\%, 4-m 18\%, 10-m 7\%, sub-mm + mm
3\%, radio 4\%, HST 15\%, other space telescopes 44\% ($\Sigma$ = 100\%).
}
\label{fig:fig4}
\end{figure*}

	The relationship between 1995-8 citation count and telescope diameter
for optical telescopes is shown in Fig. 2.
A comparison between 1991-4 and 1995-8 counts, for all telescopes,
is shown in Fig. 3.
The fractional contribution of different types of telescope to the totals,
for each period, is shown in Fig. 4.
The ground-based
optical telescopes can conveniently be divided into four classes:
1-m (diameter $<$ 2 m);
2-m (2.0 $\le$ diameter $<$ 3.0 m);
4-m (3.0 $\le$ diameter $<$ 7 m, mostly 3.5 - 4.2 m);
and 
10-m (mirror diameter $\ge$ 7 m, i.e. Keck I/II, the only ones
in this class commissioned by 1996).

Of the 14 
4-m-class optical telescopes
(i.e. excluding IRTF, UKIRT), 8 have similar 1995-8 impacts,
between 0.55 and 0.95\% of citations.
They are the AAT 4-m, CTIO 4-m, ESO 4-m, Hale 5-m (50 years old),
Lick 3-m (40 years old), 
KPNO 4-m, 
MMT 6-m and WHT 4-m telescopes.
We confirm Abt's (1985) finding that the impacts of
privately run telescopes (Hale, Lick) are not significantly
different from those of national facilities.
The CFHT 4-m has particularly high impact, with 1.8\% of citations.
The WHT ranks second.
The CFHT citations are to a total of 18 papers, 10 with first authors
at Canadian institutions, 6 US, 1 French, and 1 UK.
The Calar Alto 4-m yields a factor of $\sim$ 4 fewer citations 
than the average.
The ARC and WIYN 4-m telescopes have low impact,
perhaps 
because they were not commissioned until 1993, 1994,
and citations have yet to appear in the literature.
The other telescopes were all in regular use by 1990
(MMT was in its multi-mirror incarnation throughout this period).
No papers from the Mt. Pastukhov 
6-m telescope appear in the citations database.
In Fig. 3, $\sim$ 0.2 
of the scatter in log10 is due to small-number statistics,
but the impact of several telescopes 
has changed significantly
between 1991-4 and 1995-8 (Figs. 5, 6).


\begin{figure*}[t]
\centerline{\vbox to 11cm{
\psfig{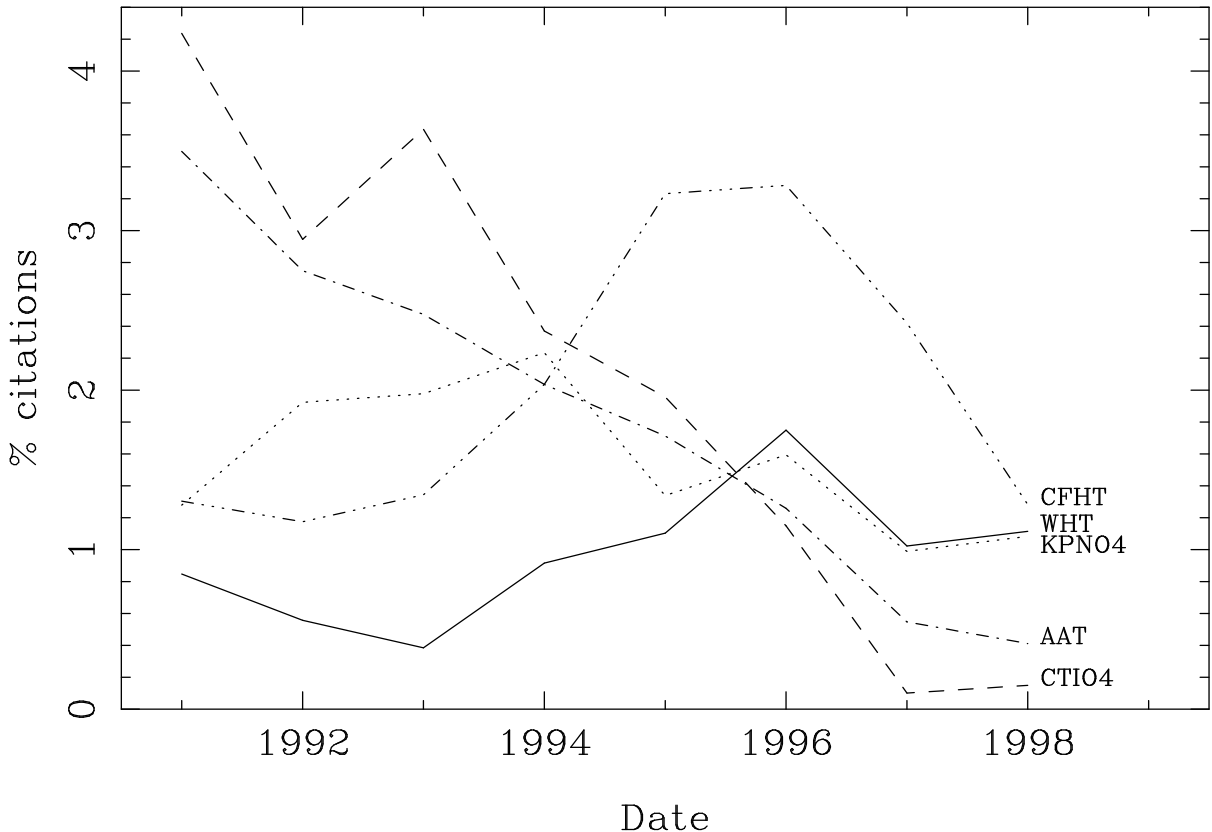}
}}
\caption{
Citations fraction (3-year running mean) vs time 
for five 4-m telescopes.  All were commissioned in the 1970s, except
the WHT (1987).
In this figure and Fig. 6 (only),
the fractions are of all observational papers, i.e. excluding theoretical
papers, to eliminate the effects of year-by-year fluctuations 
in the number of citations to the latter.
The steady decline in citation fraction for most 4-m telescopes 
reflects a corresponding rise in citations to HST and Keck I/II.
}
\label{fig:fig5}
\end{figure*}

\begin{figure*}[t]
\centerline{\vbox to 11cm{
\psfig{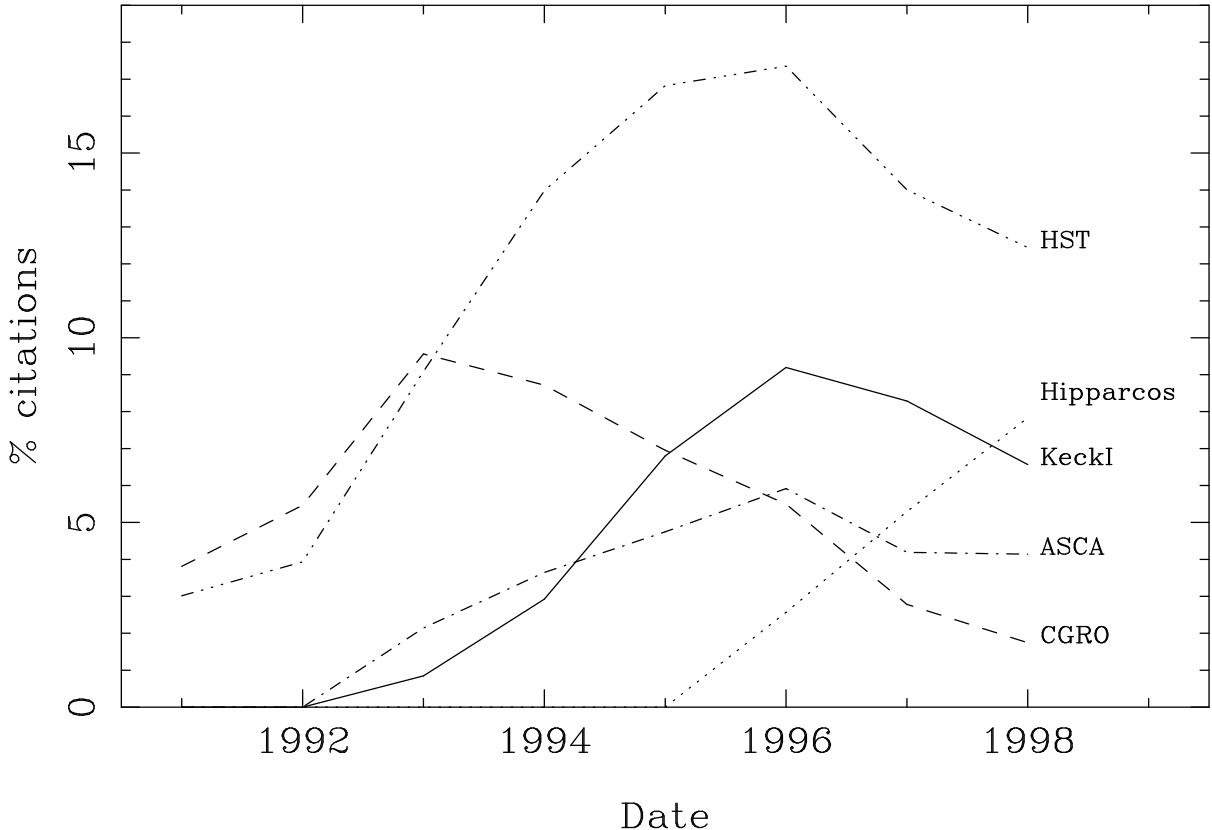}
}}
\caption{
As Fig. 5, but for
ASCA, CGRO, Hipparcos, HST and Keck I (and with different vertical scale).
After 1996, the satellites BeppoSAX, Hipparcos and RXTE receive an 
increasing fraction of the citations.
}
\label{fig:fig6}
\end{figure*}

The Keck I 10-m telescope, 
which has been in use since 1993, has an impact of 
6\% of all 1995-8 citations, 8 times that
of typical 4-m telescopes.  This factor is similar to the ratio of
collecting areas.
The (noisier) statistics for Keck II are consistent with
similar impact per unit time.
29 of the 34 cited Keck papers are on extragalactic topics.
No papers from
other 8-m - 10-m telescopes appear in the citations database (most were
commissioned after 1998).

The mean citation impact of 2-m class telescopes
is a factor $\sim$ 4 lower than that of 4-m 
telescopes, again reflecting the ratio of collecting areas.
Comparison of individual telescopes is difficult because the
numbers involved are small (Table 1).

Little can be said about the mean impact of 1-m telescopes because
of poor statistics,
but the distribution in Fig. 2 is consistent with mean impact
being roughly proportional to collecting area.
Citation counts may underestimate the 
scientific usefulness of small telescopes
because they
often provide uncited survey material on which are based 
discoveries (e.g. of high-redshift objects)
made by large telescopes.

The 1-m and 2-m telescopes together contributed one half
as much science (4.4\%) as the 4-m telescopes (8.5\%) 
during 1995-8 (the ratio was similar in 1991-4; Fig 4).
Three of the top 40 most-cited papers are based on data
from 1-m telescopes only, including two from the microlensing survey
carried out with the Mt. Stromlo 1.3-m (hence its high impact,
Figs. 2, 3).  Extrasolar planets were also discovered using a small
telescope: the Haute Provence 1.9-m (Mayor \& Queloz 1995, the 9th
most-cited paper of 1995).

The capital costs of telescopes of primary mirror diameter 
$D$ scale approximately as $D^{2.6}$, for $D <$ 7 m 
(Schmidt-Kaler \& Rucks, 1997).
8-m - 10-m telescopes currently
cost about 1/3 as much as predicted by this scaling.
Running costs dominate total cost, and scale as a similar 
(slightly smaller) power
of $D$ as do capital costs
(Abt 1980, although he only explored the relation for $D <$ 2.1 m).
The year-2000 capital costs of 2-m (median size 2.5 m), 
4-m and 10-m class telescopes are $\sim$
$\$$5M, 18M and 80M, while the citation rates are in the approximate
ratio 1:3:25 (Fig. 2). 
Therefore 2-m telescopes are roughly as
cost-effective as 4-m telescopes.
Keck I (10 m) is twice as cost-effective, but is the first of its size,
and may have a bigger scientific impact than
any of the 11 8-m - 10-m telescopes now coming online:
4 VLT, 2 Gemini, Subaru, LBT, HET, SALT and Grantecan.

The citation counts for the UKIRT 3.8-m and IRTF 3.0-m
IR telescopes are a factor $\sim$ 3 lower than
for the ground-based 4-m optical telescopes, perhaps reflecting a smaller
user community (fewer citers), see later discussion of Fig. 7.

The sub-mm telescope JCMT generated 1.3\% of 1995-8 citations,
nearly twice as much as a typical 4-m,
due largely to 6 1997/8 papers
based on data from SCUBA, the innovative sub-mm imager.
No papers from ESO's SEST appear in the citations database.
The mm telescope IRAM is highly productive, with 0.8\%
of citations.

The VLA generated 1.5\%
of 1995-8 citations.
It cost $\sim$ $\$$80M in the 1970s.
All other radio telescopes combined generated 
0.6\% of 1995-8 citations (but 1.4\% in 1991-4).  
Radio telescopes are
more prominent in the counts of Nature papers (see discussion below
of Fig. 7).
No papers from the ATCA, Effelsberg, MOST or WSRT radio telescopes
appear in the citations database. 
The VLBI observations were typically made with $\sim$ 10 - 20 telescopes
across the globe, but the contributing telescopes are often not 
identified in the published papers.

HST, launched in 1990,
generates 15 times as many citations as a 4-m telescope (11\% of the
1995-8 total),
but cost $\sim$ 100 times as much, $\sim$ $\$$2000M (a 
lot more, if the cost of servicing missions is taken into account).
ASCA, BeppoSax, CGRO, COBE, Hipparcos, ROSAT all yielded citation
fractions between 2.6 and 3.8\% of the total, 
$\sim$ 4 times higher than a 4-m telescope,
but cost $\sim$ 15 - 30 times as much.
Comparison of the cost-effectiveness of ground-based and space 
telescopes is not straightforward.  
Some space telescopes (e.g. COBE, Hipparcos)
are launched to solve a specific scientific problem which can't be tackled 
from the ground and they may have a short-lived community of citers,
so it's not clear that citation counts are a fair measure of scientific
impact.
Others, such as HST, compete more directly with ground-based
facilities (particularly now, with the advent of adaptive optics),
and can be used to tackle similar problems, so a
citation-based comparison of cost-effectiveness is fairer.
In HST's case, the cost per citation is probably $\sim$ 10 - 20 times
that for a ground-based optical telescope.
The citation shares of ASCA, CGRO, COBE and HST peaked respectively
3, 2, 3 and 6 years after launch.

In Fig. 7 the citation counts are compared with counts of papers in Nature.
Most of the telescopes contributing more than one paper to the top-cited
1000 have also generated Nature papers.
There is a positive correlation between citation fraction and Nature 
counts, and a line of slope 1 is shown.
However the constant of proportionality is different for different kinds
of telescope.  The line shows the approximate 
relation for ground-based optical
telescopes.  The ratio between citation fraction 
and number of Nature papers 
is $\sim$ twice as high for space telescopes as for optical
telescopes.  
For radio telescopes, the ratio is {\it lower} than 
for optical
telescopes.  For the VLA, the ratio is a factor of $\sim$ 3 lower, 
while for ATCA, Effelsberg, MOST, Parkes, VLBA, VLBI and WSRT, the
factor is $>\sim$ 10.
This is not due to a regional bias in publishing practice, because
the counts for optical telescopes from all countries cluster closely
about the line in Fig. 7.
If one assumes that publication in Nature is a reliable measure
of scientific worth, this result indicates that citation counts
undervalue radio work relative to optical work
(alternatively, radio work may be over-represented in Nature).
This highlights the risk of incurring metric-specific biases when
comparing the scientific impacts of telescopes of different kinds.

For IR, sub-mm and mm telescopes
(UKIRT, IRTF, JCMT, IRAM, Nobeyama, Owens Valley), 
the relation between citation fraction
and Nature counts lies between those of optical and radio telescopes.

In terms of Nature papers, 
the most productive 4-m class optical telescope during 1989-98 was
the WHT (11.5 papers), followed by the AAT (8.5).

Counts of mentions in
Sky \& Telescope and counts of papers in Nature are proportional
(with a tighter correlation than that of Fig. 7),
suggesting that the 
editors of
Nature and of Sky \& Telescope agree on
what makes interesting science.
\begin{figure*}[t]
\centerline{\vbox to 11cm{
\psfig{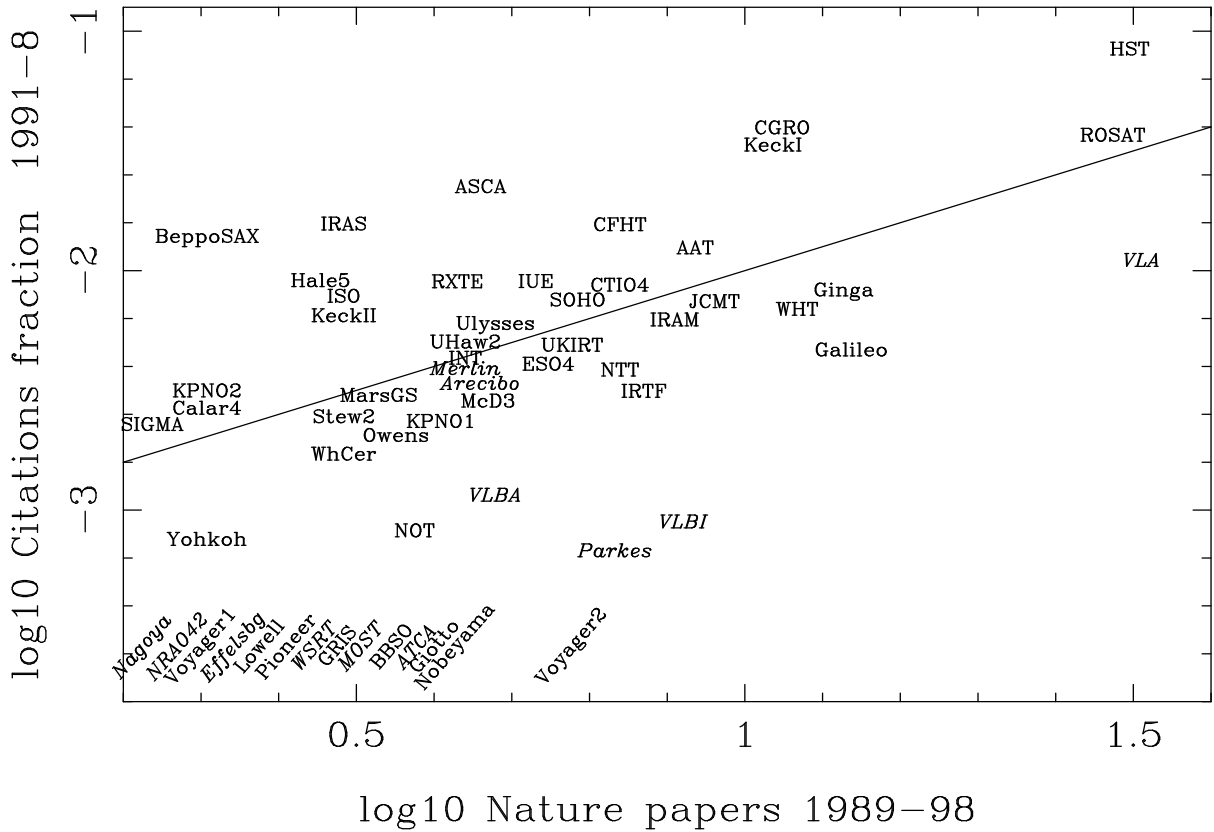}
}}
\caption{
Citations fraction 1991-8 vs number of papers in Nature,
all telescopes.
The names of long-wave radio telescopes are italicised.
The straight line has slope 1.
The telescopes appearing along the bottom edge of the diagram 
contributed data to none of the top 1000 papers, but did generate
Nature papers.
}
\label{fig:fig7}
\end{figure*}


The citations database can also be used to break down scientific impact
by region (of institution hosting first author), subject and journal.
61\% of the citations to the 1000 most-cited papers are to papers
with first authors at US institutions, 11\% UK, 20\% European (non-UK)
and 8\% other (mainly Australia, Canada, Japan).
A detailed breakdown by country is reported elsewhere
(S\'{a}nchez \& Benn, in preparation).
Breakdowns by subject and by journal are given below.

\subsection{Citation counts by subject}
Counts of citations to the 1000 most-cited papers, 
broken down by subject and date, are shown in Table 5.  
Papers on hot stars, galaxies and cosmology account for most of the
citations to papers in the top 1000.  
There was  
a significant decline of interest in AGN between
1991-4 and 1995-8.
80\% of the solar-system papers (excluding those on the sun)
were published from the USA.
Regionally, citations to work on
cool stars, hot stars and our galaxy, each break down approximately as:
62\% to papers from the USA, 
2\% UK and 36\% other countries.
For papers on galaxies, the split is 56\% USA, 21\% UK, 23\% other.
For papers on AGN and cosmology the split is
65\% USA, 12\% UK, 23\% other.
The USA is relatively strongest in cosmology, the UK in studies 
of galaxies,
other countries (mainly Europe) in studies of cool stars.

For the 4-m telescopes whose users are predominantly from North America,
98 of the 134 cited papers
(73\%) are on extragalactic topics.
For the UK (WHT, AAT), the fraction is 28 out of 33 (85\%).
At the 4-m telescopes whose users are predominantly from continental Europe,
i.e. the Calar Alto 3.5-m, NTT 3.5-m and ESO 3.6-m, 
only 8 of the 18 cited papers (44\%) are extragalactic.
These regional (or cultural) 
differences in preferred subject area are also reflected in the
extragalactic fraction in different journals:
62\% for ApJ,
83\% for MNRAS,
30\% for A\&A,
47\% for Nature.
If galactic and extragalactic astronomers had different citation
practices, these differences could
induce a regional bias in the citation counts.
To test this, we counted the number of references in 25 of the papers 
in the database from each of the US, extragalactic and European (non-UK),
stellar communities.  
The extragalactic papers had between 37 and 79 references (range excluding
smallest and largest 1/6, i.e. approximating $\pm$ 1 standard deviation for
a gaussian), the galactic ones 24 - 78.
This suggests that any difference in citing rate between the two communities
is small, and that the regional difference of subject area won't
bias the citation count.

The subject breakdown for Nature is broadly similar to that
given in Table 5, except that a larger fraction ($\approx$ 20\%)
is devoted to solar-system work.

\subsection{Citation counts by journal}
Counts of citations to the 1000 most-cited papers,
broken down by journal 
and date, are shown in Table 6.
The US journals (ApJ/S, AJ, PASP) receive 60\% of the citations,
MNRAS 9\%, A\&A/S 10\%, Nature 10\%, Science 2\%.
Most of the journals publish papers on a broad range of astronomical
themes, but
22 of the 26 papers from Science are on solar-system work (nearly all
based on data from space probes).
Most of the technical papers are published in ApJS, A\&AS or PASP.


The citation shares of most of the journals have not changed significantly
with time.
The ApJS share 
dropped from 10 to 2\% between 1991 and 1998.
Nature's share of the top 10 papers each year was 2\% in 1984-9
(1 paper), 20\% in 1990-8 (17 papers).  
Since there are $\approx$ 50 space and astronomy papers published
each year in Nature, 4\% of them are amongst the top 10 most-cited.
About 5\% of the citations
in Annual Reviews of Astronomy and Astrophysics are to Nature.

\section{Conclusions}
The scientific impacts of telescopes worldwide have been compared
on the basis of their contribution to (a) the 
1000 most-cited astronomy/space papers published 1991-8
(citation fraction), and (b) the 452 astronomy/space
papers published in Nature 1989-98 (paper count).
The conclusions below are subject to the caveats about citation bias
made in the introduction.
\newline
$\bullet$ 
During the period 1991-8, approximately 5\% of the citations 
were
generated by 1-m and 2-m ground-based optical 
telescopes, 10\% by 4-m, 4\% by Keck I/II, 
4\%  by sub-mm and radio, 8\%  by HST and 
23\% by other space telescopes.
Most of the remainder were to theoretical papers and reviews.
The strong showing by small optical telescopes 
suggests that 
cutting-edge science doesn't always require the largest aperture
available, and this augurs well for
the continued scientific impact of 4-m telescopes in the era of 8-m
telescopes.
\newline
$\bullet$
Most 4-m optical telescopes have 
citation-fraction 
and Nature-paper impacts within 30\% of the median for 
telescopes of this size.
For 1995-8, 
CFHT is the most-cited (and joint-third most prominent in Nature).
WHT is the most prominent in Nature (and the second most-cited).
\newline
$\bullet$
Citation counts correlate with counts of papers published in Nature,
but the ratio is a factor 3 
smaller for radio telescopes than for optical
telescopes, suggesting either that the former are under-cited
relative to scientific impact, or that radio work is over-represented
in the pages of Nature.
\newline
$\bullet$ 
The citation impact of ground-based optical telescopes is
proportional to collecting area and approximately proportional to 
capital cost.  Space telescopes are less cost-effective in terms 
of citation counts.
\newline
$\bullet$
There is a strong correlation between counts of papers in Nature and counts
of mentions of a telescope in Sky \& Telescope.
\newline
$\bullet$ 
61\% of citations are to papers with first authors at
US institutions, 11\% UK, 20\% European, 8\% other.
\newline
$\bullet$ 
52\% of citations are to extragalactic papers, 34\%  
stellar/galactic, 7\% solar-system, 7\% technical.
At 4-m telescopes whose users are predominantly from North America or the 
UK, 75\% of the cited papers are on extragalactics topics.
For European (non-UK) telescopes, the fraction is 44\%.
The citations to ApJ are 62\% extragalactic; to
MNRAS 83\%; to A\&A 30\%.
\newline
$\bullet$
The shares of the citation count by journal are:
ApJ 44\%, 
MNRAS 9\%, 
A\&A/S 10\%, 
Nature 11\%, 
others 29\%.

\vspace{3mm}
{\bf Acknowledgments}\\
The scans of Nature and Sky \& Telescope were made by 
1999 ING summer students
Ed Hawkins, Samantha Rix and Dan Bramich.
We are grateful to the ING and IAC librarians, Javier Mendez
and Monique Gomez, for their help, to Paul Rees
for advice on telescope costs, and to Rene Rutten, Danny Lennon
and Dennis Crabtree (the referee)
for helpful suggestions.


\clearpage




\begin{deluxetable}{lrrrrrrrrrrrr}
\tabletypesize{\scriptsize}
\tablewidth{0pt}
\tablecaption{Scientific impact of ground-based optical and IR telescopes}
\tablehead{
Telescope&Date&Diameter
&&\multicolumn{3}{c}{$\Sigma$Papers}
&&\multicolumn{3}{c}{$\Sigma$Citations \%}
&Nature&Sky\&Tel\\
&& (m)
&& 1991-4&1995-8&1991-8&&1991-4&1995-8&1991-8&1989-98&1989-98\\}
\startdata
\multicolumn{13}{l}{\bf 10-m class:} \\
KeckI  &1993    &   9.8  &&   1.0  &  25.1  &  26.1  &&   0.36 &   5.96 &   3.16 &  10.7  &  10.8  \\
KeckII &1996    &   9.8  &&   0.0  &   2.8  &   2.8  &&   0.00 &   1.44 &   0.72 &   3.0  &   0.0  \\
\multicolumn{13}{l}{\bf 4-m class:} \\
MMT    &1979    &   6.5  &&   7.4  &   3.8  &  11.2  &&   1.26 &   0.56 &   0.91 &   1.3  &   1.5  \\
Bolshoi&1975    &   6.0  &&   0.0  &   0.0  &   0.0  &&   0.00 &   0.00 &   0.00 &   1.0  &   0.0  \\
Hale5  &1948    &   5.1  &&   6.5  &   3.3  &   9.8  &&   1.03 &   0.61 &   0.82 &   2.8  &   5.5  \\
WHT    &1987    &   4.2  &&   1.6  &   3.9  &   5.5  &&   0.36 &   0.94 &   0.65 &  11.5  &   8.3  \\
CTIO4  &1976    &   4.0  &&   4.9  &   3.0  &   7.9  &&   0.90 &   0.64 &   0.77 &   6.8  &   9.3  \\
AAT    &1975    &   3.9  &&   8.3  &   3.5  &  11.8  &&   1.72 &   0.63 &   1.18 &   8.5  &   7.3  \\
KPNO4  &1973    &   3.8  &&   7.2  &   4.5  &  11.7  &&   1.22 &   0.82 &   1.02 &   0.8  &   6.8  \\
UKIRT  &1978    &   3.8  &&   3.1  &   2.0  &   5.1  &&   0.56 &   0.26 &   0.41 &   5.9  &  10.3  \\
CFHT   &1979    &   3.6  &&   4.6  &   8.4  &  13.0  &&   0.96 &   1.84 &   1.40 &   6.8  &  17.1  \\
ESO4   &1977    &   3.6  &&   1.0  &   2.5  &   3.5  &&   0.20 &   0.57 &   0.38 &   5.5  &   8.6  \\
Calar4 &1984    &   3.5  &&   2.6  &   0.2  &   2.7  &&   0.42 &   0.05 &   0.23 &   2.0  &   1.3  \\
NTT    &1989    &   3.5  &&   2.0  &   1.3  &   3.3  &&   0.26 &   0.34 &   0.30 &   6.8  &   8.8  \\
ARC    &1993    &   3.5  &&   0.0  &   1.0  &   1.0  &&   0.00 &   0.19 &   0.10 &   0.0  &   0.0  \\
WIYN   &1994    &   3.5  &&   0.0  &   0.0  &   0.0  &&   0.00 &   0.00 &   0.00 &   0.7  &   1.0  \\
Lick3  &1959    &   3.1  &&   7.0  &   4.4  &  11.4  &&   1.07 &   0.85 &   0.96 &   0.5  &   0.5  \\
IRTF   &1979    &   3.0  &&   2.4  &   1.2  &   3.6  &&   0.39 &   0.20 &   0.30 &   7.3  &   9.5  \\
\multicolumn{13}{l}{\bf 2-m class:} \\
McD3   &1969    &   2.7  &&   0.8  &   0.3  &   1.1  &&   0.49 &   0.05 &   0.27 &   4.6  &   3.0  \\
NOT    &1989    &   2.6  &&   0.0  &   0.6  &   0.6  &&   0.00 &   0.15 &   0.08 &   3.7  &   3.0  \\
DuPont &1976    &   2.5  &&   0.6  &   2.7  &   3.3  &&   0.11 &   0.48 &   0.30 &   0.0  &   0.0  \\
INT    &1984    &   2.5  &&   2.8  &   0.9  &   3.7  &&   0.58 &   0.20 &   0.39 &   4.3  &   4.8  \\
MtW3   &1917    &   2.5  &&   0.0  &   0.8  &   0.8  &&   0.00 &   0.13 &   0.07 &   0.5  &   1.0  \\
Hiltner&1986    &   2.3  &&   0.0  &   0.0  &   0.0  &&   0.00 &   0.00 &   0.00 &   1.5  &   0.0  \\
Stew2  &1969    &   2.3  &&   2.6  &   0.0  &   2.6  &&   0.46 &   0.00 &   0.23 &   3.0  &   3.5  \\
MtStr2 &1984    &   2.3  &&   0.7  &   1.1  &   1.8  &&   0.26 &   0.16 &   0.21 &   0.8  &   0.0  \\
ESO2   &1984    &   2.2  &&   0.5  &   0.0  &   0.5  &&   0.10 &   0.00 &   0.05 &   1.3  &   7.7  \\
UHaw2  &1970    &   2.2  &&   4.5  &   2.0  &   6.5  &&   0.65 &   0.30 &   0.48 &   4.3  &   6.2  \\
KPNO2  &1964    &   2.0  &&   1.3  &   1.9  &   3.1  &&   0.23 &   0.27 &   0.25 &   2.0  &   7.0  \\
\multicolumn{13}{l}{\bf 1-m class:} \\
Haute2 &1958    &   1.9  &&   0.5  &   1.0  &   1.5  &&   0.22 &   0.42 &   0.32 &   0.0  &   0.0  \\
Whipple    &1970    &   1.5  &&   1.0  &   1.0  &   2.0  &&   0.20 &   0.14 &   0.17 &   0.8  &   2.0  \\
CTIO2  &1968    &   1.5  &&   0.0  &   1.1  &   1.1  &&   0.00 &   0.19 &   0.10 &   0.3  &   1.2  \\
Pal2   &1908    &   1.5  &&   0.0  &   0.0  &   0.0  &&   0.00 &   0.00 &   0.00 &   1.3  &   0.0  \\
MtStr1 &1941    &   1.3  &&   1.0  &   3.0  &   4.0  &&   0.55 &   0.80 &   0.68 &   1.0  &   0.0  \\
UOWO   &        &   1.3  &&   0.0  &   0.0  &   0.0  &&   0.00 &   0.00 &   0.00 &   1.3  &   0.0  \\
LC1    &        &   1.0  &&   2.0  &   1.4  &   3.4  &&   0.52 &   0.23 &   0.38 &   0.0  &   0.0  \\
Marly  &        &   1.0  &&   0.0  &   1.0  &   1.0  &&   0.00 &   0.19 &   0.09 &   0.0  &   0.0  \\
CTIO1  &1967    &   0.9  &&   2.0  &   0.5  &   2.5  &&   1.18 &   0.10 &   0.64 &   1.3  &   0.5  \\
KPNO1  &1960    &   0.9  &&   2.9  &   0.2  &   3.2  &&   0.41 &   0.04 &   0.22 &   4.0  &   5.4  \\
Lowell &        &   0.5  &&   0.0  &   0.0  &   0.0  &&   0.00 &   0.00 &   0.00 &   2.3  &   0.0  \\
PalSch &1948    &   1.2  &&   0.0  &   1.1  &   1.2  &&   0.01 &   0.23 &   0.12 &   0.0  &   0.6  \\
UKST   &1973    &   1.2  &&   0.6  &   0.8  &   1.3  &&   0.09 &   0.11 &   0.10 &   1.5  &   4.3  \\
\multicolumn{13}{l}{\bf Solar:} \\
BBSO   &1969    &        &&   0.0  &   0.0  &   0.0  &&   0.00 &   0.00 &   0.00 &   3.0  &   1.0  \\
\multicolumn{13}{l}{} \\
Theoretical  &        &        && 160.3  & 114.4  & 274.6  &&  31.79 &  21.84 &  26.82 &     &    \\
Reviews &        &        &&  38.0  &  18.0  &  56.0  &&   7.91 &   3.53 &   5.72 &     &     \\
\enddata
\tablecomments{The telescope abbreviations are those used in the
figures, and are explained in Table 4.
Column 2 gives the date of commissioning.
Columns 4 - 6 refer to counts of papers (sums of fractions)
from the 1000 most-cited, and columns 7 - 9 to citations of these
papers.
}
\end{deluxetable}

\clearpage

\begin{deluxetable}{lrrrrrrrrrrrr}
\tabletypesize{\scriptsize}
\tablewidth{0pt}
\tablecaption{Scientific impact of ground-based sub-mm, 
radio and cosmic-ray telescopes}
\tablehead{
Telescope&Date&
&&\multicolumn{3}{c}{$\Sigma$Papers}
&&\multicolumn{3}{c}{$\Sigma$Citations \%}
&Nature&Sky\&Tel\\
&& 
&& 1991-4&1995-8&1991-8&&1991-4&1995-8&1991-8&1989-98&1989-98\\
}
\startdata
\multicolumn{13}{l}{\bf Sub-mm and mm:} \\
CSO    &1988    &        &&   1.4  &   0.5  &   1.9  &&   0.27 &   0.07 &   0.17 &   1.0  &   1.0  \\
JCMT   &1987    &        &&   0.2  &   6.0  &   6.2  &&   0.06 &   1.34 &   0.70 &   9.0  &   4.0  \\
SEST   &1987    &        &&   0.0  &   0.0  &   0.0  &&   0.00 &   0.00 &   0.00 &   0.5  &   0.0  \\
IRAM   &1979    &        &&   3.2  &   4.6  &   7.8  &&   0.54 &   0.78 &   0.66 &   8.0  &   1.0  \\
Nobeyama&1989    &        &&   0.0  &   0.0  &   0.0  &&   0.00 &   0.00 &   0.00 &   3.3  &   0.0  \\
Owens  &        &        &&   1.0  &   2.0  &   3.0  &&   0.16 &   0.28 &   0.22 &   3.5  &   1.0  \\
\multicolumn{13}{l}{\bf Radio:} \\
Arecibo&1963    &        &&   3.0  &   1.0  &   4.0  &&   0.48 &   0.16 &   0.32 &   4.5  &   6.4  \\
ATCA   &1988    &        &&   0.0  &   0.0  &   0.0  &&   0.00 &   0.00 &   0.00 &   3.3  &   1.0  \\
Effelsberg&1972    &        &&   0.0  &   0.0  &   0.0  &&   0.00 &   0.00 &   0.00 &   2.2  &   2.0  \\
GB91   &1978    &        &&   2.0  &   0.0  &   2.0  &&   0.40 &   0.00 &   0.20 &   0.5  &   0.0  \\
Merlin &1990    &        &&   3.0  &   0.0  &   3.0  &&   0.74 &   0.00 &   0.37 &   4.0  &   5.0  \\
MOST   &1967    &        &&   0.0  &   0.0  &   0.0  &&   0.00 &   0.00 &   0.00 &   3.0  &   1.0  \\
Nagoya &        &        &&   0.0  &   0.0  &   0.0  &&   0.00 &   0.00 &   0.00 &   1.8  &   0.0  \\
NRAO42 &1965    &        &&   0.0  &   0.0  &   0.0  &&   0.00 &   0.00 &   0.00 &   2.0  &   0.0  \\
Parkes &1961    &        &&   1.0  &   0.0  &   1.0  &&   0.13 &   0.00 &   0.06 &   6.7  &   0.0  \\
VLA    &1980    &        &&   2.4  &   5.6  &   8.1  &&   0.55 &   1.54 &   1.04 &  31.9  &  33.4  \\
VLBA   &1993    &        &&   0.0  &   1.2  &   1.2  &&   0.00 &   0.22 &   0.11 &   4.7  &   2.8  \\
VLBI   &        &        &&   0.0  &   1.2  &   1.2  &&   0.00 &   0.17 &   0.08 &   8.2  &   2.5  \\
WSRT   &1970    &        &&   0.0  &   0.0  &   0.0  &&   0.00 &   0.00 &   0.00 &   2.2  &   1.8  \\
\multicolumn{13}{l}{\bf Cosmic-ray:} \\
FlysEye&        &        &&   1.0  &   1.0  &   2.0  &&   0.16 &   0.17 &   0.16 &   0.0  &   0.0  \\
WhCer  &1992    &        &&   1.0  &   1.0  &   2.1  &&   0.18 &   0.14 &   0.16 &   3.0  &   2.0  \\
\enddata
\tablecomments{
The telescope abbreviations are those used in the
figures, and are explained in Table 4.
Column 2 gives the date of commissioning.}
\end{deluxetable}

\clearpage

\begin{deluxetable}{lrrrrrrrrrrrr}
\tabletypesize{\scriptsize}
\tablewidth{0pt}
\tablecaption{Scientific impact of space (and airborne) telescopes}
\tablehead{Telescope&Date&
&&\multicolumn{3}{c}{$\Sigma$Papers}
&&\multicolumn{3}{c}{$\Sigma$Citations \%}
&Nature&Sky\&Tel\\
&& 
&& 1991-4&1995-8&1991-8&&1991-4&1995-8&1991-8&1989-98&1989-98\\}
\startdata
\multicolumn{13}{l}{\bf $\gamma$-ray:} \\
BeppoSAX&1996    &        &&   0.0  &   9.5  &   9.5  &&   0.00 &   2.63 &   1.31 &   2.0  &   0.5  \\
CGRO   &1991    &        &&  19.4  &  14.7  &  34.1  &&   4.04 &   2.93 &   3.48 &  11.0  &  14.0  \\
GRIS   &1988    &        &&   0.0  &   0.0  &   0.0  &&   0.00 &   0.00 &   0.00 &   3.0  &   0.0  \\
\multicolumn{13}{l}{\bf X-ray:} \\
ASCA   &1993    &        &&   2.5  &  16.5  &  19.0  &&   0.90 &   3.33 &   2.12 &   4.5  &   3.8  \\
Einstein&1978    &        &&   4.9  &   1.3  &   6.2  &&   1.01 &   0.31 &   0.66 &   1.0  &   1.7  \\
EXOSAT &1983    &        &&   4.0  &   0.0  &   4.0  &&   0.81 &   0.00 &   0.41 &   1.0  &   1.5  \\
Ginga  &1987    &        &&   6.0  &   1.5  &   7.5  &&   1.10 &   0.23 &   0.67 &  13.2  &   1.3  \\
HXT    &1991    &        &&   1.5  &   0.0  &   1.5  &&   0.23 &   0.00 &   0.11 &   0.0  &   0.0  \\
ROSAT  &1990    &        &&  18.7  &  18.3  &  37.0  &&   3.79 &   3.15 &   3.47 &  29.3  &  16.8  \\
RXTE   &1995    &        &&   0.0  &  10.6  &  10.6  &&   0.00 &   1.70 &   0.85 &   4.2  &   4.3  \\
SIGMA  &1989    &        &&   3.0  &   0.0  &   3.0  &&   0.43 &   0.00 &   0.22 &   1.7  &   2.3  \\
SXT    &1991    &        &&   1.5  &   0.0  &   1.5  &&   0.53 &   0.00 &   0.26 &   0.0  &   0.0  \\
Yohkoh &1991    &        &&   0.0  &   1.0  &   1.0  &&   0.00 &   0.14 &   0.07 &   2.0  &   0.5  \\
\multicolumn{13}{l}{\bf Optical/UV:} \\
HUT    &1990    &        &&   0.0  &   1.6  &   1.6  &&   0.00 &   0.38 &   0.19 &   1.5  &   0.0  \\
Hipparcos&1989    &        &&   0.0  &  13.0  &  13.0  &&   0.00 &   3.14 &   1.57 &   1.0  &   2.0  \\
HST    &1990    &        &&  25.5  &  55.9  &  81.5  &&   4.58 &  11.33 &   7.95 &  30.8  &  72.1  \\
IUE    &1978    &        &&   6.1  &   2.5  &   8.7  &&   1.21 &   0.49 &   0.85 &   5.3  &   2.8  \\
KAO    & 1995   &        &&   0.0  &   0.0  &   0.0  &&   0.00 &   0.00 &   0.00 &   1.5  &   2.0  \\
\multicolumn{13}{l}{\bf IR:} \\
IRAS   &1983    &        &&  10.3  &   4.3  &  14.6  &&   2.04 &   0.92 &   1.48 &   3.0  &  10.7  \\
ISO    &1995    &        &&   0.0  &   7.1  &   7.1  &&   0.00 &   1.48 &   0.74 &   3.0  &   1.5  \\
\multicolumn{13}{l}{\bf Radio:} \\
COBE   &1989    &        &&  16.4  &  12.9  &  29.3  &&   4.64 &   3.34 &   3.99 &   1.0  &   7.5  \\
\multicolumn{13}{l}{\bf Solar:} \\
SMM    &1980    &        &&   2.0  &   0.0  &   2.0  &&   0.29 &   0.00 &   0.15 &   1.0  &   1.0  \\
SOHO   &1995    &        &&   0.0  &   8.8  &   8.8  &&   0.00 &   1.42 &   0.71 &   6.0  &   3.5  \\
SUMER  &        &        &&   0.0  &   1.3  &   1.3  &&   0.00 &   0.25 &   0.13 &   0.0  &   0.0  \\
Ulysses&1990    &        &&   4.0  &   2.0  &   6.0  &&   0.67 &   0.30 &   0.48 &   4.7  &   3.0  \\
\multicolumn{13}{l}{\bf Planetary missions:} \\
Galileo&1995    &        &&   0.0  &   5.0  &   5.0  &&   0.00 &   0.88 &   0.44 &  13.5  &   5.0  \\
Giotto &1985    &        &&   0.0  &   0.0  &   0.0  &&   0.00 &   0.00 &   0.00 &   4.0  &   0.0  \\
Magellan&1990    &        &&   2.0  &   0.0  &   2.0  &&   0.27 &   0.00 &   0.13 &   0.0  &   0.0  \\
MarsGS &1998    &        &&   0.0  &   0.0  &   0.0  &&   0.00 &   0.00 &   0.00 &   4.0  &   0.0  \\
Pioneer&1972    &        &&   0.0  &   0.0  &   0.0  &&   0.00 &   0.00 &   0.00 &   2.3  &   0.0  \\
Voyager1 &1977    &        &&   0.0  &   0.0  &   0.0  &&   0.00 &   0.00 &   0.00 &   2.0  &   3.0  \\
Voyager2 &1977    &        &&   0.0  &   0.0  &   0.0  &&   0.00 &   0.00 &   0.00 &   6.0  &   4.0  \\
\enddata
\tablecomments{
The telescope abbreviations are explained in Table 4.
Column 2 gives the date of launch (or of arrival at planet).}
\end{deluxetable}

\clearpage

\begin{deluxetable}{ll}
\tablewidth{0pt}
\tablecaption{Telescope abbreviations (as used in the figures)}
\tablehead{
Abbrev.&Telescope, and location (if ground-based)\\}
\startdata
AAT    &Anglo-Australian Telescope, Siding Spring                                       \\
Apollo & Apollo solar-wind mission \\
ARC    &Advanced Research Consortium telescope, Apache Point, New Mexico                \\
Arecibo&Arecibo radio telescope, Puerto Rico                                                 \\
ASCA   &Advanced X-ray Satellite for Cosmology \& Astrophysics                                     \\
ATCA & Australia Telescope Compact Array (radio), Narrabri\\
ATT-horn & AT\&T Bell Labs 20-ft horn antenna, New Jersey \\
BBSO & Big Bear Solar Observatory, California \\
BeppoSAX&BeppoSAX $\gamma$-ray telescope                                                    \\
Bolshoi&Mt Pastukhov telescope, Georgia                                                           \\
Calar4  &Calar Alto 3.5-m telescope                                                               \\
CFHT   &Canada-France-Hawaii Telescope, Mauna Kea                                       \\
CGRO   &Compton Gamma Ray Observatory                                                   \\
COBE   &Cosmic microwave Background Explorer                                            \\
CSO    &Caltech Sub-mm Observatory, Mauna Kea                                   \\
CTIO1  &0.9-m  telescope, Cerro Tololo                                                          \\
CTIO2  &1.5-m  telescope, Cerro Tololo                                                            \\
CTIO4  &Blanco 4-m telescope, Cerro Tololo                                                   \\
DuPont &du Pont 2.5-m telescope, Las Campanas                                                     \\
Effelsberg & Effelsberg 100-m radio telescope, Germany \\
Einstein&Einstein X-ray telescope                                                        \\
ESO2   &ESO 2.2-m telescope, La Silla                                                             \\
ESO4   &ESO 3.5-m telescope, La Silla                                                             \\
EXOSAT &X-ray telescope                                                                          \\
FlysEye&Fly's Eye air-shower detector, Utah                                             \\
Galileo&Galileo mission to Jupiter                                                      \\
GB91   &Green Bank 91-m radio telescope                                                 \\
Gemini & Two 8.1-m telescopes, Mauna Kea and Cerro Pachon \\
Ginga  &X-ray telescope                                                           \\
Giotto & Giotto mission to comet P/Halley \\
Grantecan & Gran Telescopio Canario (10-m), La Palma\\
GRIS & Gamma-Ray Imaging Spectrometer (balloon-borne) \\
Hale5   &Hale Telescope, Mt Palomar, California                                          \\
Haute2 &Haute Provence 1.9-m  telescope                                                          \\
HET & Hobby-Eberly Telescope, Davis Mountains, Texas \\
Hiltner & Hiltner telescope, Kitt Peak \\
Hipparcos&Hipparcos astrometry mission                                                                      \\
HST    &Hubble Space Telescope                                                          \\
HUT    &Hopkins UV Telescope (on the Space Shuttle)                          \\
HXT    &Hard X-Ray Telescope                                                            \\
INT    &Isaac Newton Telescope, La Palma                                                \\
IRAM   &French mm telescope, Pico Veleta, Granada                           \\
IRAS   &IR Astronomy Satellite                                                          \\
IRTF   &NASA IR Telescope Facility, Mauna Kea                                           \\
ISO    &IR Space Observatory                                                            \\
IUE    &International UV Explorer                                                          \\
JCMT   &James Clark Maxwell Telescope, Mauna Kea                                        \\
KAO & Kuiper Airborne Observatory \\
KeckI  &Keck I, Mauna Kea                                                               \\
KeckII &Keck II, Mauna Kea                                                              \\
KPNO1  &0.9-m telescope, Kitt Peak, Arizona                                                                            \\
KPNO2  &2-m telescope, Kitt Peak, Arizona                                                              \\
KPNO4  &Mayall 4-m telescope, Kitt Peak, Arizona                                            \\
LBT & Large Binocular Telescope (2 8.4-m), Mt Graham, Arizona \\
LC1    &1-m telescope, Las Campanas                                                               \\
Lick3  &Shane 3-m telescope, Mt Hamilton, California                             \\
Lowell & Lowell Observatory 0.5-m telescope, Arizona \\
Magellan&Magellan mission to Venus                                                       \\
Marly & Marly 1-m telescope, La Silla \\
MarsGS & Mars Global Surveyor \\
McD3   &McDonald 2.7-m telescope, Davis Mountains, Texas                                                \\
Merlin &Merlin radio interferometer, UK                                                  \\
MMT    &Multiple-Mirror Telescope, Mt Hopkins, Arizona                                  \\
MOST & Molonglo Observatory Synthesis Telescope (radio), Australia \\
MtStr1 &1-m telescope, Mt Stromlo                                                                      \\
MtStr2 &2.3-m telescope, Mt Stromlo                                                 \\
MtW2 & 1.5-m telescope, Mt Wilson, California\\
MtW3 & Hooker 2.5-m telescope, Mt Wilson, California\\
Nagoya & University of Nagoya 4-m radio telescope (solar), Japan \\
Nobeyama & Nobeyama mm array, Japan \\
NOT    &Nordic Optical Telescope, La Palma                                              \\
NRAO42 & NRAO 42-m radio telescope, Green Bank, West Virginia \\
NTT    &New Technology Telescope, La Silla                                              \\
Owens  &Owens Valley mm array, Big Pine, California                                                 \\
Pal2   &60-in telescope, Mt Palomar, California                                    \\
PalSch   &Oschin 48-in Schmidt Telescope, Mt Palomar, California                                    \\
Parkes & Parkes 64-m radio telescope, Narrabri \\
Pioneer & Pioneer 10, 11 space probes \\
ROSAT  &Roentgen X-ray Satellite                                                        \\
RXTE   &Rossi X-ray Timing Explorer                                                   \\
SALT & South African Large Telescope (modelled on HET) \\
SEST & Swedish / ESO Sub-mm Telescope \\
SIGMA  &X-ray telescope       \\
SMM    &Solar Maximum Mission                                                           \\
SOHO   &Solar and Heliospheric Observatory                                              \\
Stew2   &Steward 2.3-m (Bok) telescope, Kitt Peak, Arizona                                              \\
Subaru & Japanese 8.3-m telescope, Mauna Kea \\
SUMER  &UV telescope on SOHO                                                          \\
SXT    &Soft X-ray Telescope                                                            \\
UHaw2    &University of Hawaii 88-inch, Mauna Kea                                         \\
UKIRT  &UK Infrared Telescope, Mauna Kea                                                \\
UKST   &UK Schmidt Telescope, Siding Spring                                             \\
Ulysses&Ulysses Solar Satellite                                                         \\
UOWO & 1.3-m telescope, University of Western Ontario \\
VLA    &Very Large Array radio telescope, New Mexico                                    \\
VLBA   &Very Long Baseline Array, USA                                                   \\
VLBI   &Very Long Baseline Interferometry                                               \\
VLT & Very Large Telescope (4 8.2-m telescopes), Mt Paranal\\
Voyager 1,2 & Voyager missions to outer solar system \\
WhCer    &Whipple Cerenkov telescope, Mt Hopkins, Arizona                                                      \\
WHT    &William Herschel Telescope, La Palma                                            \\
WIYN   &Wisconsin/Indiana/Yale/NOAO telescope, Kitt Peak, Arizona                       \\
WSRT & Westerbork Synthesis Radio Telescope, Netherlands \\
Whipple    &1.5-m Whipple Telescope, Mt Hopkins, Arizona                                                                             \\
Yale & Yale 1.0-m telescope, Cerro Tololo \\
Yohkoh & X-ray telescope\\
\enddata
\end{deluxetable}

\clearpage
\begin{deluxetable}{lrrrrrrrrr}
\tablewidth{0pt}
\tablecaption{Citation \% by subject (including theoretical papers)}
\tablehead{
Subject & 1991& 1992 & 1993&1994&1995&1996&1997&1998&Total\\}
\startdata
Technical   &    6.7&   11.7&    5.0&    4.9&    6.4&    6.7&    7.5&    4.9&    6.7\\
Sun         &    8.4&    1.1&    3.1&    2.4&    4.9&    2.4&    3.4&    1.8&    3.4\\
Solar System   &    3.5&    4.7&    3.7&    1.6&    2.0&    2.5&    2.7&    5.7&    3.3\\
Exoplanets  &    0.0&    0.0&    0.0&    0.0&    1.7&    2.3&    1.9&    0.0&    0.7\\
Cool stars  &    8.5&   14.8&   12.2&   12.5&    5.9&    9.3&    4.9&    3.5&    9.0\\
Hot stars   &    7.4&   12.2&   20.2&   14.8&   12.7&    8.0&   19.6&   23.8&   14.8\\
Our galaxy      &   11.5&    5.8&   10.2&   11.0&    6.8&    7.9&   11.9&    7.9&    9.1\\
Galaxies    &   20.8&   21.7&   14.9&   16.1&   21.1&   31.5&   19.4&   24.1&   21.2\\
AGN         &   21.3&    7.1&   15.0&   12.5&   21.4&    6.2&    8.7&    3.0&   11.9\\
Cosmology   &   11.9&   20.9&   15.8&   22.8&   16.5&   22.7&   20.1&   24.3&   19.4\\
\enddata
\end{deluxetable}

\clearpage

\begin{deluxetable}{lrrrrrrrrr}
\tablecaption{Citation \% by journal (including theoretical papers)}
\tablewidth{0pt}
\tablehead{
Journal & 1991& 1992 & 1993&1994&1995&1996&1997&1998&Total\\}
\startdata
ApJ   &   48.6&   40.4&   39.0&   36.4&   37.0&   48.8&   49.8&   55.0&   44.4\\
Nature   &    7.6&    9.1&    6.9&   13.8&   13.0&    8.3&   13.7&   13.7&   10.7\\
MNRAS    &    9.7&    7.5&   11.0&    7.9&    6.7&   11.1&    9.3&    7.6&    8.8\\
ApJS  &    7.4&   13.2&    8.7&   13.6&   12.2&    3.9&    2.6&    1.8&    7.9\\
A\&A    &    7.5&    5.3&   11.5&    2.9&    2.7&    7.2&    3.0&    9.8&    6.3\\
AJ    &    3.3&    7.1&    5.0&    3.6&    7.6&    6.4&    5.1&    4.1&    5.3\\
ARA\&A  &    3.4&    5.7&    6.3&    6.5&    2.6&    4.6&    2.0&    0.6&    4.0\\
A\&AS  &    0.9&    7.1&    4.6&    3.1&    0.0&    1.2&    8.3&    0.8&    3.3\\
Science  &    0.0&    0.0&    2.1&    0.5&    3.6&    5.3&    2.7&    3.7&    2.2\\
PASP  &    0.9&    1.9&    1.9&    2.8&    7.1&    1.2&    0.0&    0.6&    2.1\\
Other &   10.8&    2.8&    2.9&    8.4&    7.6&    2.0&    3.5&    2.3&    5.0\\
\enddata
\end{deluxetable}

\end{document}